\documentclass{ifacconf}

\usepackage{graphicx}      
\usepackage{natbib}        

\usepackage{amssymb}
\usepackage{amsmath} 
\usepackage{soul,color}

\usepackage{tikz}
\usepackage{booktabs}
\usepackage{nicematrix}
\usepackage{subcaption}
\usepackage{epstopdf}
\usepackage{stfloats}

\usepackage{mleftright} 
\mleftright

\usepackage{xcolor}

\usetikzlibrary{arrows.meta,positioning,calc}
\usepackage[most]{tcolorbox}

\newcommand \dRelgz	{ k_\mathrm{g} }
\newcommand \Relcz		{ \mathcal R_\mathrm{c0} }
\newcommand \Relgz		{ \mathcal R_\mathrm{g0}  }

\newcommand \lambdasat	{ \lambda_\mathrm{sat} }
\newcommand \param		{ \theta } 
\newcommand \parid {\hat \param} 

\newcommand \Rel			{ \mathcal R }
\newcommand \g {\mathrm{g}}
\newcommand \tf			{ t_\mathrm{f} }

\newcommand \vcoil		{ u }
\newcommand \uff {\vcoil_{f{\!}f}}
\newcommand \vel			{ v }
\newcommand \vc {v_\mathrm{c}}

\newcommand \yff {\hat{y}_2}

\newcommand \zmax			{ z_\mathrm{max} }
\newcommand \zmin			{ z_\mathrm{min} }

\newcommand \x {x}
\newcommand \f {f}
\newcommand \G {x}
\newcommand \um {u}
\newcommand \h {h}
\newcommand \IO {IO}

\newcommand{\hide}[1]{}

\NewDocumentCommand{\MiFiguraContent}{}{
\begin{figure*}[!t]
    \centering
	\def\sumoffset{1mm}
	\def\sumoffsetaux{1.5mm}
    \def\arrowoffset{3mm}
	\def\nodex{10mm}
	\def\nodey{12mm}
	\def\arrowsep{3mm}
	\def\lwt{0.3mm}
	\def\lwn{0.4mm}
	\def\lwd{0.5mm}
    \hspace{-12mm}
    \begin{subfigure}{0.45\linewidth}
        \centering
	\begin{tikzpicture}[font=\small,
		node distance = \nodey and \nodex,
		box/.style = {draw, minimum height=6mm, minimum width=12mm, align=center},
		bigbox/.style = {draw, dashed,draw=linen, minimum height=13mm, minimum width=24mm, align=center},
		sum/.style = {circle, draw, node contents={}},
		>={Stealth[width=2mm,length=3mm]}
		]
        \newlength{\distsum}
        \pgfmathsetlength{\distsum}{1.5*\nodex}
		\node (ref) [] {};
        \node (0ref)[above=of ref.center, xshift=\arrowoffset, yshift=-\nodey] {$y_1^\mathrm{ref}(t)$};
		\node (ff) [box, right=of ref, yshift=0] {Feedforward \\controller};
 		\coordinate[right=of ff,xshift=0] (c0);
        \node (plant) [box, minimum height=2\baselineskip, right=of ff,xshift=0.65*\nodex] {System};
        \node (z_out) [right=of plant.center, yshift=\arrowoffset, xshift=\nodex] {}; 
        \node (est) [box, below=of plant.center,yshift=0, anchor=center] {Open-loop\\predictor};
        \coordinate[below=of c0.center] (c1);
		\node (s1) [sum, right=of est.center ,xshift=\distsum, yshift=-\arrowoffset, anchor=center];
        \node (s1+) [above right=of s1.center, xshift=-\nodex, yshift=-\nodey+\sumoffset, scale=0.75] {\textbf{$\mathbf +$}};
		\node (s1-) [above left=of s1.center, xshift=\nodex-\sumoffsetaux, yshift=-\nodey, scale=0.75] {$\mathbf -$};
        \coordinate[right=of plant.center, xshift=\distsum, yshift=-\arrowoffset] (c2);
        \node (cost) [box, below=of s1.center,yshift=0.25*\nodey, anchor=center] {Cost};		
        \node (opt) [box, below=of est.center, yshift=-\arrowoffset+0.25*\nodey, anchor=center] {Iterative\\adaptation law};
        \node (xpre) [right=of est.center, yshift=\arrowoffset, xshift=\nodex] {};
        \coordinate[below=of ff.center, yshift=-\arrowoffset] (c3);
        \coordinate[below=of ff.center, yshift=-\nodey-\arrowoffset+0.25*\nodey] (c4);
        
        \draw[->,line width=\lwt] (ref) -- (ff);
        \draw[->,line width=\lwt] (ff) -- node[above,xshift=-0.5*\arrowoffset] {$\uff(t,\parid_{k})$} (plant);
        \draw[->,line width=\lwt] (c1) -- (est);
        \draw[->,line width=\lwt] ([yshift=\arrowoffset]plant.east) -- node[above, xshift=0.5*\nodex-0.4*\arrowoffset]{$y_1(t)$} (z_out);
        \draw[-,line width=\lwt] ([yshift=-\arrowoffset]plant.east) -- node[above, yshift=-0.7mm, xshift=-1.2*\arrowoffset] {$y_2(t)$} (c2);
        \draw[->,line width=\lwt] (c2) -- (s1);
        \draw[-,line width=\lwt] (c0) -- (c1);]
        \draw[->,line width=\lwt] ([yshift=-\arrowoffset]est.east) -- node[above, yshift=-0.7mm, xshift=-0.5*\arrowoffset] {$\yff(t,\parid_{k})$} (s1);
        \coordinate[right=of est.center ,xshift=\distsum-\arrowoffset/3, yshift=\arrowoffset] (c5);
        \coordinate[right=of est.center ,xshift=\distsum+\arrowoffset/3, yshift=\arrowoffset] (c6);
        \coordinate (c55) at ($ (c5) !.5! (c6) $); 
        \draw[->,line width=\lwt] ([yshift=\arrowoffset]est.east) -- node[above, xshift=0.5*\nodex] {$\hat y_1(t,\parid_k)$} (xpre);
    
        \draw[->,line width=\lwn] (s1) -- node[right] {$e_2(t)$} (cost);
        \draw[dashed,->,line width=\lwn] (cost) -- node[above] {$J_{k}$} (opt);
        \draw[dashed,line width=\lwn] (opt) -- node[above] {$\parid_{k+1}$} (c4);
        \draw[dashed,line width=\lwn] (c4) -- (c3);
        \draw[dashed,->,line width=\lwn] (c3) -- (ff);
        \draw[dashed,->,line width=\lwn] (c3) -- ([yshift=-\arrowoffset]est.west);
        \filldraw [black] (c0) circle (1.5pt);
        \filldraw [black] (c3) circle (1.5pt);
        
	\end{tikzpicture}
    \caption{Conceptual idea of the proposed control architecture}
        \label{fig:control0}
    \end{subfigure}
    \hspace{3mm}
    \begin{subfigure}{0.45\linewidth}
        \centering
	\begin{tikzpicture}[font=\small,
		node distance = \nodey and \nodex,
		box/.style = {draw, minimum height=6mm, minimum width=12mm, align=center},
		bigbox/.style = {draw, dashed,draw=linen, minimum height=13mm, minimum width=24mm, align=center},
		sum/.style = {circle, draw, node contents={}},
		>={Stealth[width=2mm,length=3mm]}
		]
        \pgfmathsetlength{\distsum}{1.5*\nodex}
		\node (ref) [] {};
        \node (0ref)[above=of ref.center, xshift=\arrowoffset, yshift=-\nodey] {$y_1^\mathrm{ref}(t)$};
		\node (ff) [box, right=of ref, yshift=0] {Flatness-based \\controller};
 		\coordinate (c0) at ($ (ref) !.5! (ff.west) $); 
		\node (plant) [box, minimum height=2\baselineskip, right=of ff,xshift=0.65*\nodex] {System};
        \node (z_out) [right=of plant.center, yshift=\arrowoffset, xshift=\nodex] {}; 
        \node (est) [box, below=of plant.center,yshift=0, anchor=center] {Flatness-based\\predictor};
        \coordinate[below=of c0.center] (c1);
		\node (s1) [sum, right=of est.center ,xshift=1.15*\distsum, yshift=-\arrowoffset, anchor=center];
        \node (s1+) [above right=of s1.center, xshift=-\nodex, yshift=-\nodey+\sumoffset, scale=0.75] {\textbf{$\mathbf +$}};
		\node (s1-) [above left=of s1.center, xshift=\nodex-\sumoffsetaux, yshift=-\nodey, scale=0.75] {$\mathbf -$};
        \coordinate[right=of plant.center, xshift=1.15*\distsum, yshift=-\arrowoffset] (c2);
        \node (cost) [box, below=of s1.center,yshift=0.25*\nodey, anchor=center] {Cost};		
        \node (opt) [box, below=of est.center, yshift=-\arrowoffset+0.25*\nodey, anchor=center] {Iterative\\adaptation law};
        \node (xpre) [right=of est.center, yshift=\arrowoffset, xshift=\nodex] {};
        \coordinate[below=of ff.center, yshift=-\arrowoffset] (c3);
        \coordinate[below=of ff.center, yshift=-\nodey-\arrowoffset+0.25*\nodey] (c4);
        
        \draw[->,line width=\lwt] (ref) -- (ff);
        \draw[->,line width=\lwt] (ff) -- node[above,xshift=-0.5*\arrowoffset] {$\uff(t,\parid_{k})$} (plant);
        \draw[->,line width=\lwt] ([yshift=\arrowoffset]plant.east) -- node[above, xshift=0.5*\nodex+0.1*\arrowoffset]{$y_1(t)$} (z_out);
        \draw[-,line width=\lwt] ([yshift=-\arrowoffset]plant.east) -- node[above, yshift=-0.7mm, xshift=-0.65*\arrowoffset] {$y_2(t)$} (c2);
        \draw[->,line width=\lwt] (c2) -- (s1);
        \draw[-,line width=\lwt] (c0) -- (c1);
        \coordinate[below=of ff.center ,xshift=-\arrowoffset/3] (c5);
        \coordinate[below=of ff.center ,xshift=\arrowoffset/3] (c6);
        \coordinate (c55) at ($ (c5) !.5! (c6) $); 
        \draw[-,line width=\lwt] (c1) -- (c5);
        \draw[-,line width=\lwt] (c5) to[out=90, in=180] ([yshift=\arrowoffset/3]c55) to[out=0,  in=90] (c6);
        \draw[->,line width=\lwt] (c6) -- (est);
        
        \draw[->,line width=\lwt] ([yshift=-\arrowoffset]est.east) -- node[above, yshift=-0.7mm, xshift=-0.4*\arrowoffset] {$\yff(t,\parid_{k})$} (s1);
        
        \draw[->,line width=\lwt] ([yshift=\arrowoffset]est.east) -- node[above, xshift=0.5*\nodex] {$\hat y_1(t,\parid_k)$} (xpre);
    
        \draw[->,line width=\lwn] (s1) -- node[right] {$e_2(t)$} (cost);
        \draw[dashed,->,line width=\lwn] (cost) -- node[above] {$J_{k}$} (opt);
        \draw[dashed,line width=\lwn] (opt) -- node[above] {$\parid_{k+1}$} (c4);
        
        \draw[dashed,line width=\lwn] (c4) -- (c3);
        \draw[dashed,->,line width=\lwn] (c3) -- (ff);
        \draw[dashed,->,line width=\lwn] (c3) -- ([yshift=-\arrowoffset]est.west);
        \filldraw [black] (c0) circle (1.5pt);
        \filldraw [black] (c3) circle (1.5pt);
        
	\end{tikzpicture}
        \caption{Efficient flatness-based control alternative}
        \label{fig:control}
    \end{subfigure}
    \vspace{-2mm}

    \caption{Control diagram. The subscript $k$ denotes the variables of the $k$-th evaluation of the R2R adaptation law. The objective is to control the non-measurable output $y_1$ by adapting the model parameters $\parid$ through the prediction error of the measurable output $y_2$}
	\label{fig:v_ctrl_diag}
\end{figure*}

}

\begin{document}
\begin{frontmatter}

\title{
Run-to-Run Indirect Trajectory Tracking Control of Electromechanical Systems Based on Identifiable and Flat Models\thanksref{footnoteinfo}} 


\thanks[footnoteinfo]{
This work was supported in part via projects PID2024-159279OB-I00 and CPP2024-011615, funded by \mbox{MICIU/AEI/10.13039/501100011033} and by ERDF/EU, and in part by the Gobierno de Aragón under Project DGA T45{\_}23R}

\author[First]{Eloy Serrano-Seco}
\author[First]{Edgar Ramirez-Laboreo}
\author[First]{Eduardo Moya-Lasheras}

\address[First]{Departamento de Informatica e Ingenieria de Sistemas (DIIS) and Instituto de Investigacion en Ingenieria de Aragon (I3A),\\Universidad de Zaragoza, 50018 Zaragoza, Spain,\\ (e-mail:
\{eserranoseco,
ramirlab, 
emoya\}%
@unizar.es%
)}

\begin{abstract}                
Differentially flat models are frequently used to design feedforward controllers for electromechanical systems. However, control performance depends on model accuracy, which makes feedback imperative. This paper presents a control scheme for electromechanical systems in which measuring or estimating the output to be controlled---typically the position---is not feasible. It employs an identifiable-model-based controller and predictor, coupled with an iterative loop that updates model parameters using the error between a measurable output and its prediction. Simulations on electromechanical switching devices show effective tracking of the desired position trajectory using only coil current measurements.
\end{abstract}


\end{frontmatter}

\section{Introduction}
Differential flatness is a structural property shared by a large class of electromechanical system models~\citep{rigatos2015nonlinear}. In a differentially flat $n$th order system, the $n$th derivative of the output is the first one in which the input appears explicitly~\citep{levine2011}.
This feature enables the calculation of the equivalent system input and the corresponding state variables from a desired output through model inversion.
One of the most notable applications is the design of continuous-time feedforward controllers (also known as exact feedforward linearization), which avoids linear approximations and therefore preserves accuracy.
In addition, feedforward controllers achieve superior performance in terms of response time and tracking accuracy compared to other control strategies. Due to all these advantages, flatness-based feedforward controllers have been proposed to control the motion of a wide variety of electromechanical systems. These include cranes~\citep{lobe2018flatness}, micro-opto-electro-mechanical systems actuators~\citep{li2024determination}, electrical drives~\citep{ghadbane2024energy}, electrohydraulic systems~\citep{sarkar2024novel}, and quadrotors~\citep{sun2022comparative}, among others. 

Feedforward controllers are, however, highly sensitive to modeling and parameter identification errors. For this reason, they are typically combined with feedback mechanisms to improve robustness and compensate for model inaccuracies. Examples include classical Proportional--Integral feedback controllers~\citep{sarkar2024novel}, optimal control~\citep{lobe2018flatness}, Gaussian-process online model learning~\citep{tesfazgi2023model}, and adaptive parameter-adjustment methods~\citep{fu2022adaptive}. 

Unfortunately, in some scenarios, feeding back the variable to be controlled, i.e., the position, is not technically or economically feasible. An example is that of commercial electromechanical switching devices, in which the long-standing control problem of soft landing needs to be solved without the possibility of measuring or estimating the position of the movable component.
Recent works~\citep{moya2023IFAC} exploit the repetitive operation of such devices by applying run-to-run (R2R) control~\citep{sachs1991R2Rjournal}. 
This technique iteratively adjusts the parameters of the control process between successive runs to achieve the desired output quality.

In some of our previous works \citep{serrano2022,edgar2024ECC}, we employed an R2R control scheme with a feedforward controller in the main loop and an iterative adaptation law in the outer loop. The iterative adaptation law is supplied with a measurement related to the impact energy (impact velocity, impact sound, or the bounces produced) without considering the full motion trajectory. 
Although effective, this approach does not guarantee that the desired trajectory is accurately followed. This limitation is particularly relevant when there are multiple control objectives~\citep{serrano2022}, or stricter reliability and repeatability are required.

This paper introduces a new control structure that leverages model identifiability~\citep{vajda1989similarity} to guarantee accurate position tracking, even when the tracked variable cannot be measured. Building upon the scheme proposed in~\cite{moya2023IFAC}, we incorporate a model-based predictor that estimates a measurable output—--different from the controlled variable--—using the same input as the feedforward controller. Both the controller and the predictor are obtained through model inversion of an identifiable flat model, and are adapted cycle by cycle by an external loop. The contribution is focused on the tracking error of the variable of interest: the unified predictor–controller architecture that provides information on the tracking error based on predicted and measured output, the redefined input of the iterative adaptation law that allows for the indirect minimization of the tracking error, and the use of a structurally identifiable model that provides a theoretical guarantee of the tracking.
Simulation results validate the proposed approach, demonstrating that it solves the soft-landing problem as effectively as state-of-the-art methods while ensuring accurate reference tracking. Furthermore, the practical identifiability implications are analyzed in light of the obtained results.

\MiFiguraContent
\section{Problem formulation} \label{sec:problem}
Let us consider a generic nonlinear system $\Sigma$, whose dynamics can be described using a state-space model,
\begin{equation} \label{eq:system}
     \Sigma \left(\param\right):\begin{cases}
\dot \x = \f(\x,\um, \param) \\
y = \h(\x, \param)
\end{cases} \hspace{-4mm},
\end{equation}
where $\x$, $\um$, and $y$ are the state, input, and output vectors, respectively. The functions $\f$ and $\h$ also depend on the parameter vector $\param$.

The system presents the particularity that the output to be controlled, from now on denoted as $y_1$, is non-measurable, but another output, from now on denoted as $y_2$, is measurable.
For the purpose of analyzing specific subsets of outputs, we define the subsystems $\Sigma_1$ and $\Sigma_2$ as the systems obtained from $\Sigma$ by selecting $y_1$ and $y_2$ as the output vectors, respectively,
\begin{equation} \label{eq:subsystem}
\hspace{-1mm}
     \Sigma_1\left(\param\right):\begin{cases}
\dot \x = \f(\x,\um, \param) \\
y_1 = \h_1(\x, \param)
\end{cases} \hspace{-4mm}, \hspace{-1mm}\quad\qquad
\Sigma_2\left(\param\right):\begin{cases}
\dot \x = \f(\x,\um, \param) \\
y_2 = \h_2(\x, \param)
\end{cases} \hspace{-4mm}.
\end{equation}
The subsystems are characterized by $\Sigma_1$ being \textit{differentially flat} and $\Sigma_2$ being \textit{structurally identifiable}.

The control objective is for the output $y_1$ to track a reference signal using only the measurement of $y_2$. The proposed control scheme, which exploits the repetitive operation of many electromechanical systems, aims to solve this problem under uncertainty in the value of $\param$.

\section{Control} \label{sec:control}

The conceptual idea of the controller is schematized in Fig.~\ref{fig:control0}. It includes a feedforward term that computes the input signal $\uff$ based on the desired reference signal $y_1^\mathrm{ref}$ of the non-measurable output.
Meanwhile, a predictor calculates the expected measurable output $\yff$.
The input signal is applied to the electromechanical system to make it follow the predefined trajectory, and the prediction error $e_2 = y_2 - \hat y_2$ is used to calculate a performance index $J$. 
Finally, the run-to-run adaptation law updates the model parameters $\parid$ involved in both the feedforward controller and the predictor to reduce $e_2$ in subsequent cycles.

\subsection{Open-loop control: Flatness-based control blocks} \label{subs:Open-Loop}
The conceptual idea in Fig.~\ref{fig:control0} requires numerical integration to compute the predicted output $\yff$.
The use of the flatness property, however, results in the more computationally efficient control scheme depicted in Fig.~\ref{fig:control}.

A feature of flat systems is that the states $\x$ and inputs $\um$ can be expressed as functions of the input $y$ and a finite number of its derivatives,
\begin{equation} \label{eq:isflat}
\Sigma_1 \,\, \mathrm{is \, flat} \;\Longleftrightarrow\;
\begin{cases}
\x = \G\left(y_1,\dot{y}_1,\ddot{y}_1,...,\param\right) \\
\um = \um\left(y_1,\dot{y}_1,\ddot{y}_1,...,\param\right)
\end{cases}\hspace{-4mm}.
\end{equation}
This enables the straightforward design of feedforward controllers by means of model inversion. The system input to be applied can be computed by substituting $y_1$ and its derivatives with the desired output $y_1^\mathrm{ref}$ and its derivatives,
\begin{equation}
    \uff(t,\parid) = \um\left(y_1^\mathrm{ref},\dot{y}_1^\mathrm{ref},\ddot{y}_1^\mathrm{ref},...,\parid\right).
\end{equation}

On the other hand, a predictor is also proposed. Given \eqref{eq:subsystem} and \eqref{eq:isflat}, the flatness property also enables computing the predicted output vector $\hat{y}$ from $y_1^\mathrm{ref}$ and its derivatives,
\begin{equation} \label{eq:predictions}
    \begin{aligned}
        \hat{y}_1(t,\parid) = \h_1\left(\G\left(y_1^\mathrm{ref},\dot{y}_1^\mathrm{ref},\ddot{y}_1^\mathrm{ref},...,\parid\right), \parid\right), \\
        \hat{y}_2(t,\parid) = \h_2\left(\G\left(y_1^\mathrm{ref},\dot{y}_1^\mathrm{ref},\ddot{y}_1^\mathrm{ref},...,\parid\right), \parid\right).
    \end{aligned}
\end{equation}
Note that, as the predictor is designed by model inversion,
$\hat{y}_1 = y_1^\mathrm{ref}$.

In this manner, both the controller and the predictor are defined by analytical expressions, which offers an improvement over the general idea depicted in Fig.~\ref{fig:control0} by eliminating the need for numerical integration.

\subsection{Structural identifiability: tracking guarantee} \label{subs:Ident}
In any electromechanical system, several factors lead to discrepancies between control parameters $\parid$ and system parameters $\param$, ranging from manufacturing tolerances to system wear. This reduces the controller accuracy and introduces prediction errors.
To address this, the structural identifiability of the system is analyzed to determine whether the parameters can be accurately estimated.

Structural identifiability \citep{vajda1989similarity} is a theoretical property that ensures that the model parameters can be uniquely determined from the input–output behavior under ideal conditions, i.e., noise-free measurements, no disturbances, and sufficient excitation. In other words, two distinct parameter sets must always result in different outputs for identical initial conditions and inputs.
Let $\IO_{\Sigma}(\param): \um(t) \mapsto y(t), \forall t \in [t_0, \tf]$ be the input-output map of a system $\Sigma$ in the time-domain interval from $t_0$ to $\tf$ with initial state $\x_0(\param)$. Structural identifiability implies,
\begin{equation} \label{map2}
    \IO_{\Sigma_2}(\parid) = \IO_{\Sigma_2}(\param) \, \Longleftrightarrow \, \parid = \param,
\end{equation}
where $\IO_{\Sigma_2}(\param)$ and $\IO_{\Sigma_2}(\parid)$ represent the input-output maps of the system and predictor considering the measurable output, $y_2$ and $\yff$, respectively.

Then, provided that the controlled-output function $\h_1$ also depends on the same parameter vector $\theta$~\eqref{eq:subsystem}, it follows that
\begin{equation} \label{eq:map1}
    \parid = \param \, \Longrightarrow \, \IO_{\Sigma_1}(\parid) = \IO_{\Sigma_1}(\param),
\end{equation}
where $\IO_{\Sigma_1}(\param)$ and $\IO_{\Sigma_1}(\parid)$ represent the input-output maps of the system and the predictor with $y_1$ and $\hat{y}_1$ as output, respectively.

Since all input–output maps share the same input $\uff$, and under the condition of sufficient excitation, matching the predicted and measured outputs ($\hat{y}_2(t,\parid) = y_2(t,\param), \,\, \forall t$) implies that the system parameters are correctly estimated. 
This implies that the true output equals the predicted output, which, in turn, coincides with the reference by design ($y_1(t,\param) = \hat{y}_1(t,\parid) =  y_1^\mathrm{ref}(t), \,\, \forall t$).

\subsection{Iterative feedback loop: Run-to-run adaptation} \label{subs:Adapt}
Due to the identifiability property, the predicted error $e_2$ reports discrepancies between the reference $y_1^\mathrm{ref}$ and the actual output $y_1$.
However, no mechanism is currently available to correct these discrepancies. A natural choice would be real-time feedback strategies, but these are not suitable: $e_2$ does not provide a quantitative value of the trajectory error, simultaneous state–parameter estimation is unreliable due to accumulation of initial errors, and real-time $e_2$ control invalidates the identifiability conclusions, due to, by $\IO_{\Sigma}$ definition, $e_2(t) = 0, \, \forall t\in [t_0, \tf]$ with $x_0(\parid)$ that are not known at midstream.
As an alternative, leveraging the repetitive nature of a wide variety of electromechanical systems, an R2R strategy is adopted.

The proposed R2R strategy consists of a cost-evaluation block that returns a performance index $J$, and an iterative adaptation law that updates the parameters of both the controller and the predictor to optimize $J$. A key advantage of this structure is its flexibility, as it is compatible with a wide range of algorithms adapted to online optimization available in the literature.
The index $J$ is computed from the measurable output error,
\begin{equation}\label{eq:cost_exp}
    J = \int_{t_0}^{\tf} {e_2}(t)^{\intercal} \, {e_2}(t) \, \mathrm d t,
\end{equation}
where the interval $[t_\mathrm{0}, \tf]$ is large enough to capture the expected repetitive dynamics of the system.
Therefore, by structural identifiability, minimizing the error in the measurable output theoretically identifies the system parameters, and ensures tracking of the reference signal $y_1^\mathrm{ref}$.

\section{Case study}
To demonstrate the effectiveness of the proposal, the soft-landing control problem in commercial electromechanical switching devices is considered. Uncontrolled activations of these low-cost devices cause strong impacts between fixed and movable components, leading to undesirable phenomena. This motivates the design of soft-landing control strategies for the position of a movable component that is unmeasurable.

These devices are based on a single-coil reluctance actuator. We consider a system with linear gap reluctance and magnetic saturation in the core. Following the description provided in~\cite{moya2023IFAC}, the system can be described by a state-space model as
\begin{equation}    \label{eq:dyn}
\begin{aligned}
    \dot \x &= \begin{cases}
    \dot z \,\,= \vel, \\   
    \dot \vel \,\,= \frac{1}{m}\, \left( -k_\mathrm{s} \, \left(z-z_\mathrm{s}\right) - c_\mathrm{f} \, \vel - \frac{1}{2}\,\dRelgz \,\lambda^2 \,\right), \\ 
    \dot \lambda \; = -R \, \lambda\, \left(\Relgz + \dRelgz \, z + \frac{\Rel_\mathrm{c0}}{1-|\lambda|/\lambda_\mathrm{sat}}\right) + u, 
    \end{cases} \\
    y &= \begin{cases}
    y_1 = z, \\
    y_2 = \lambda\, \left(\Relgz + \dRelgz \, z + \frac{\Rel_\mathrm{c0}}{1-|\lambda|/ \lambda_\mathrm{sat}}\right),
\end{cases}
\end{aligned}
\end{equation}
where the state variables $z$, $\vel$, and $\lambda$ are the position of the component to be controlled, its velocity, and the magnetic flux linkage, respectively. The outputs are $y_1$ the non-measurable position and $y_2$ the measurable coil current. The input $u$ is the voltage, and the remaining parameters are constants that form the parameter vector $\rho$ that defines the system,
\begin{equation}\label{eq:param_real}
        \rho = \left[
        \,\, m \,\,\,\, k_\mathrm{s} \,\,\,\, z_\mathrm{s} \,\,\,\, c  \,\,\,\, k_\mathrm{g} \,\,\,\, \Rel_\mathrm{g0} \,\,\,\, \Rel_\mathrm{c0} \,\,\,\, \lambdasat \,\,\,\, R \,\, 
    \right]^\intercal .
\end{equation}

The control proposal requires a model that is differentially flat (with $y_1$ as the output) and structurally identifiable (when $y_2$ is measured). The previous model does not satisfy the latter condition, because some parameters exhibit a structural correlation, e.g., $m$, $k_\mathrm{s}$, $c_\mathrm{f}$, and $k_\g$. To resolve this, we apply the following state transformation: $x_1 = \Relgz + \dRelgz \, z$, $x_2 = \dRelgz \, v$, $x_3 = \lambda$. This results in an identifiable alternative realization,
\begin{equation}    \label{eq:id}
\begin{aligned}
    \dot \x &= \begin{cases}
    \dot x_1 = \, x_2, \\
    \dot x_2 = -\param_1 \, x_1- \param_2 \, x_2 - \frac{1}{2}\, \param_4 \,{x_3}^2  -\param_3,  \\  
    \dot x_3 = -\param_7 \, x_3\, \left(x_1 + \frac{\param_5}{1-|x_3|/\param_6}\right) + u, 
    \end{cases} \\
    y &= \begin{cases} 
    y_1 \,= x_1, \\
    y_2 \,= \, x_3\, \left(x_1+ \frac{\param_5}{1-|x_3|/\param_6}\right),
\end{cases}
\end{aligned}
\end{equation} 
where the new parameter vector, $\param \in \mathbb R^7$, is formed by a nonlinear combination of $\rho$, $\theta = \varphi(\rho)$, with
\begin{equation}
\label{eq:parid}
\begin{array}{lll}
     \param_1 = k_\mathrm{s}/m,   &\quad&\param_2 = c_\mathrm{f}/m,  \\   
    \param_3 = -k_\mathrm{s} \, \left( k_\g \, z_\mathrm{s} + \Relgz \right) / m,   &\quad&\param_4 = k_\g^2/m, \\
    \param_5 = \Relcz,  \quad \qquad  \param_6 = \lambdasat,  &\quad&\param_7 = R .
\end{array}       
\end{equation} 

The structural identifiability property of this model can be demonstrated by means of the Local State Isomorphism theorem~\citep{vajda1989similarity}. It is also straightforward to check that the model remains flat with $y_1$ as the flat output.

\section{Simulated results}
In this section, we present results obtained by simulation. The objective is to analyze whether the control scheme is able to replicate the desired trajectory and whether the adaptation law estimates the true values of the parameters. Both premises are evaluated under the condition of solving the soft-landing problem of electromechanical switching devices, i.e., achieving that the movable component of these devices reaches the end-of-stroke with zero velocity. 

The new proposal, denoted as \mbox{R2R-IM} (R2R-Indirect Measurement), is compared to a previous approach presented in~\cite{moya2023IFAC}, denoted as \mbox{R2R-DM} (R2R-Direct Measurement). While \mbox{R2R-DM} minimizes a cost function $J$ based directly on the impact velocity, $v_c$, the proposed \mbox{R2R-IM} utilizes the indirect measurement of $y_2$ (coil current). For a fair comparison, both control schemes use the same feedforward controller and optimization method.

\subsection{Description of the simulated experiments}
\begin{table}[t]
    \begin{center}
    \renewcommand{\arraystretch}{1.1} 
    \caption{Nominal model parameter values}\label{tb:param} 
    \vspace{-2mm}
        \begin{tabular}{ccccc}
            Parameter \hspace{0pt} & Value & \hspace{6pt} &
	    	Parameter \hspace{0pt} & Value \\
	    	\cmidrule{1-2} \cmidrule{4-5}
	    	$ \Rel_\mathrm{c0} $ & $1.35 \,\mathrm{H^{-1}} $ &&
	    	$ \zmin $ & $0$ \\
	    	$ \lambdasat $ & $0.0229 \,\mathrm{Wb} $ &&
	    	$ \zmax $ & $10^{-3}\,\mathrm{m}$ \\
    		$ \Rel_\mathrm{g0}  $ & $3.88 \,\mathrm{H^{-1}} $ &&
    		$ m $ & $1.6 \times 10^{-3} \,\mathrm{kg} $ \\
    		$ k_\g $ & $7.67 \,\mathrm{H^{-1}/m} $ &&
    		$ k_\mathrm{s} $ & $ 55 \,\mathrm{N/m} $ \\
    		$ R $ & $ 50 \, \mathrm{\Omega}$ &&
            $ z_\mathrm{s} $ & $ 0.0181 \,\mathrm{m} $ \\
	    	\cmidrule{1-2} \cmidrule{4-5} 
        \end{tabular}
    \end{center}
\end{table}

In the simulations, it is assumed that the dynamics of the system is completely described by the model equations previously presented in~\eqref{eq:dyn}. 
The feedforward controllers of both strategies being compared, i.e., \mbox{R2R-DM} and \mbox{R2R-IM}, and the model-based predictor of \mbox{R2R-IM} have been designed by inversion of the model~\eqref{eq:id}, as explained in Section~\ref{subs:Open-Loop}. 
Note that while the simulated system and the control blocks utilize different representations, they represent the same underlying dynamics.
Nevertheless, it is assumed that there is some uncertainty in the parameter values initially used by the controller, as occurs in real scenarios due to manufacturing tolerances.
To emulate this, $10\,000$ devices have been generated, creating $10\,000$ different $\rho$-vectors~\eqref{eq:param_real}, varying the parameters of the real device randomly and independently, with a uniform probability distribution between $95\,\%$ and $105\,\%$ of $\rho^\mathrm{nom}$, the nominal values of $\rho$ (see Table~\ref{tb:param}). Note that the variability in the identifiable parameters $\param$ of the devices generated can exceed $\pm 5\%$, because these are combinations of $\rho$~\eqref{eq:parid}, $\param = \varphi\left(\rho^\mathrm{nom}\,(1+\epsilon)\right)$ where $\epsilon \sim \mathcal{U}(-0.05,0.05)$.

To analyze both control schemes, two Monte Carlo analyses are conducted. These analyses comprise $10\,000$ trials, with each trial performing a sequence of $600$ switching operations.
In the first operation, the control uses the nominal parameter vector, $\parid = \param^\mathrm{nom} = \varphi(\rho^\mathrm{nom})$. In subsequent operations, $\parid$ is adapted via an optimization method based on the Nelder--Mead method~\citep{nelder1965}.
The last parameter $\param_7$, i.e., the coil resistance $R$, is excluded from the adaptation law because it can be easily estimated from the voltage (input $u$) and current (measurable output $y_2$) in steady state. Note that \mbox{R2R-DM} would require an additional measurement or knowledge of the resistance.

With respect to $y_1^\mathrm{ref}$, a $7$th-degree polynomial is used to define it. The coefficients have been tuned to produce a reference similar to that reported in \cite{serrano2022}, while ensuring that they satisfy the soft-landing requirements. 
The duration of the defined $y_1^\mathrm{ref}$ has been set to 4.5\,ms, ensuring that the trajectory is feasible, i.e. compatible with the system dynamics, for all $10\,000$ different $\param$-vectors.

\subsection{Results and discussion}
\begin{figure}[t]
    \centering
    \begin{subfigure}{1\linewidth}
        \centering
    	\includegraphics{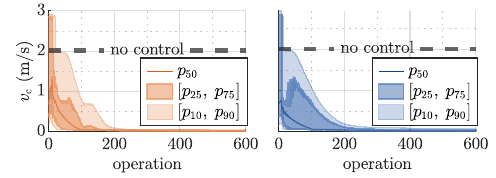} \vspace{-1mm}
        \caption{Distribution of impact velocities $\vc$}
        \label{fig:results_vc}
    \end{subfigure}
    \hfill
    \begin{subfigure}{1\linewidth}
        \centering
        \includegraphics{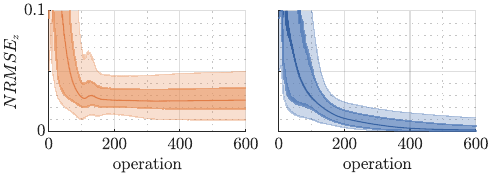} \vspace{-1mm}
        \caption{Distribution of $NRMSE_z$}
        \label{fig:results_ez}
    \end{subfigure}
    \begin{subfigure}{1\linewidth}
        \centering
        \includegraphics{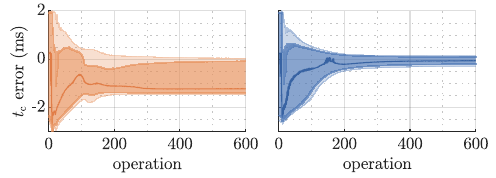} \vspace{-1mm}
        \caption{Distribution of the impact time $t_\mathrm{c}$ error}
        \label{fig:results_tc}
    \end{subfigure} \vspace{-6mm}
    \caption{Analyzed indices as a function of the number of switching operations. The graphs show the median ($p_{50}$), interquartile range ($[p_{25},\,p_{75}]$), and $10$th-–$90$th percentile interval ($[p_{10},\,p_{90}]$) from 10\,000 simulated experiments. Left: \mbox{R2R-DM}; right: proposed \mbox{R2R-IM}}
    \label{fig:results_index}
\end{figure}

\begin{figure}[t]
    \begin{center} \vspace{1mm}
    	\includegraphics{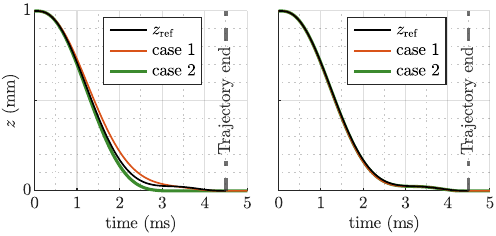} \vspace{-2mm}
        \caption{Position trajectories obtained in the last iteration of the 10\,000 simulated experiments. Only the trajectories corresponding to the worst cases of Fig.~\ref{fig:results_ez} (case 1) and Fig.~\ref{fig:results_tc} (case 2) are plotted. Left: \mbox{R2R-DM}; right: proposed \mbox{R2R-IM}}
        \label{fig:results_tray}
    \end{center}
\end{figure}

Different indices are analyzed from the control results. Firstly, since the objective of soft-landing control is to reduce the impact energy, Fig.~\ref{fig:results_vc} illustrates the distribution of impact velocities $\vc$ obtained along the operations. To highlight the control effectiveness, the graphs also include the mean cost of a conventional non-controlled switching operation (specifically, with a 30~V constant activation). 
As can be seen, even though the new proposal does not use the impact velocity for the adaptation law, similar results are obtained with both strategies. If we examine the value of the 90th percentile at iteration 600, the impact-velocity reduction only increases from $2.15\,\%$ to $3.5\,\%$ with the new proposal. However, if we examine the value of the median, the new control scheme exhibits faster initial convergence while achieving the same final value.

The second aspect to assess is whether the new proposal improves reference tracking. To facilitate a more interpretable analysis, instead of comparing $y_1$ of the identifiable model, we compare the corresponding position trajectories: the position $z$ generated by the simulated device and the reference trajectory $z_\mathrm{ref}$, which can be derived from $y_1$ and $y_1^\mathrm{ref}$.
Fig.~\ref{fig:results_ez} represents the distribution of the normalized root-mean-square position error $NRMSE_z$ obtained along the operations,
\begin{equation}
    NRMSE_z = \sqrt{
\frac{\int_{t_0}^{\tf} \left(z(t) - z_{\mathrm{ref}}(t) \right)^2 \, \mathrm{d}t
}{\int_{t_0}^{\tf} {z_\mathrm{ref}}^2(t) \, \mathrm{d}t}
}.
\end{equation}
As can be seen, R2R-IM reduces position errors further as the number of operations increases, whereas the previous approach R2R-DM converges to larger errors.

It is worth noting that although the tracking errors from both control schemes may seem comparable, the effect can be significant in certain devices. As an example, electromechanical relays should complete movement in a reduced time interval to prevent electric arcs and reduce wear~\citep{ksiazkiewicz2019change}. The instant at which the movable component contacts the end-of-stroke and does not take off again is denoted as the impact time, $t_\mathrm{c}$. The $t_\mathrm{c}$ error distribution is shown in Fig.~\ref{fig:results_tc}. As illustrated, the new control scheme considerably reduces the $t_\mathrm{c}$ error window from $1.5\,\mathrm{ms}$ to $0.4\,\mathrm{ms}$.

To illustrate the effect on the position trajectory, Fig.~\ref{fig:results_tray} displays the trajectories at the final iteration ($k=600$) corresponding to the worst-case scenarios identified in Fig.\ref{fig:results_ez} (case 1) and Fig.\ref{fig:results_tc} (case 2). 
As can be seen, \mbox{R2R-DM} reaches impact velocities close to zero, but follows a trajectory different from the reference trajectory, \(z_{\mathrm{ref}}\). This behavior can be attributed to the presence of multiple points in the search space of the iterative adaptation law that yield costs similar to the global minimum, i.e., points that achieve a soft landing.
In contrast, \mbox{R2R-IM} exploits the identifiability property to ensure that the cost function reaches zero only when \(\parid=\param\). Consequently, this approach not only achieves a soft landing but also accurately tracks the reference trajectory and lands at the desired instant.

\subsection{Parameter identification discussion}
\begin{figure}[t]
    \begin{center}
        \includegraphics{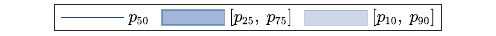} \\
    	\includegraphics{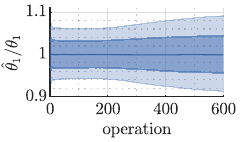} 
        \includegraphics{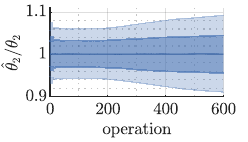}\\ 
        \includegraphics{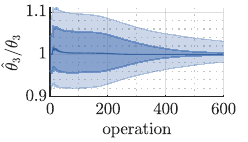} 
        \includegraphics{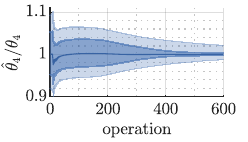}\\
        \includegraphics{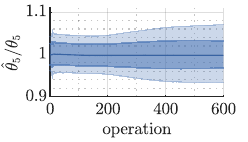} 
        \includegraphics{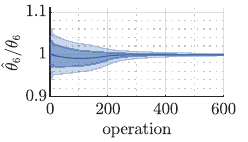}\\
        \vspace{-2mm}
        \caption{Normalized adapted-parameter values with respect to the real ones as a function of the number of switching operations. Each graph shows the median ($p_{50}$), interquartile range ($[p_{25},\,p_{75}]$), and $10$th-–$90$th percentile interval ($[p_{10},\,p_{90}]$) from 10\,000 simulated experiments using \mbox{R2R-IM}}
        \label{fig:results_param}
    \end{center}
\end{figure}

The previous subsection has demonstrated that \mbox{R2R-IM} solves the soft-landing problem as well as \mbox{R2R-DM}, while achieving trajectory tracking. However, the question remains whether it is capable of identifying the model parameters to ensure tracking.

Fig.~\ref{fig:results_param} shows the evolution of the optimized parameters of the identifiable model, $\parid$, normalized with respect to the true values, $\param$. Surprisingly, only half of the adapted parameters ($\parid_3$, $\parid_4$, and $\parid_6$) converge to the true values. To understand why the parameters are not correctly estimated despite the fact that position tracking is achieved, a sensitivity analysis is performed.

The lack of identifiability of certain parameters arises because the control scheme was developed under ideal conditions, assuming sufficient excitation for their identification. In practice, however, practical identifiability must be assessed. To analyze practical identifiability, the sensitivity of $\yff$ with respect to the control parameter vector, $\parid$, has been calculated. Additionally, the sensitivity of $y_1$ with respect to $\parid$ has been computed to analyze the effect of each parameter on the tracking performance. 

Table~\ref{tb:sensitivities} summarizes the dimensionless integrated-squared sensitivity vectors $\bar{S}^{\yff}$ and $\bar{S}^{y_1}$, where each element is calculated as follows:
\begin{equation}
    S_i^\mathcal{Y}= \param_i^2\int_{t_0}^{\tf} \left(\frac{\partial\mathcal{Y}(t)}{\partial \param_i}\right)^2 \mathrm{d}t, \quad \bar{S}_i^\mathcal{Y} = S_i^\mathcal{Y}/\max \left( S^\mathcal{Y}\right),
\end{equation}
where the superscript $\mathcal{Y}$ refers to the analyzed output.

As observed, the parameter convergence results and the sensitivity analysis lead to the same conclusion. Parameters $\parid_3$ and $\parid_4$ exhibit a significant influence on the unmeasured trajectory $y_1$, implying that parameter errors could degrade the tracking performance. However, this does not pose an issue because they also show a high sensitivity with respect to the predicted output $\yff$. Since the controller drives the measured $y_2$ to $\yff$, these parameters are reliably identified, ensuring the correct trajectory is maintained. In contrast, $\parid_6$ presents the ideal case: it has a negligible impact on $y_1$, and it can be identified because the predicted output $\yff$ shows appreciable sensitivity to this parameter. 
In essence, tracking is achieved if $y_1$ is sensitive only to control parameters that also sensitize $\yff$, which can therefore be accurately estimated.

Note also that this sensitivity analysis can be interpreted as a simplified version of the formal Fisher Information Matrix–based analysis, focusing only on the diagonal terms.
The obtained sensitivities, or the full Fisher matrices, can serve as bases for search space reduction techniques, which can improve convergence rates~\citep{edgar2024ECC}.

\begin{table} [t]
    \begin{center}
    \setlength{\tabcolsep}{2pt} 
    \newcolumntype{C}[1]{>{\centering\arraybackslash}m{#1}}
    \small
    \caption{Dimensionless integrated-squared \newline sensitivities with respect to the parameters}\label{tb:sensitivities} 
    \vspace{-1mm}
    \begin{tabular}{m{2em}|C{4em}C{4em}C{4em}C{4em}C{4em}C{4em}}
                   & $\parid_1$ & $\parid_2$ & $\parid_3$ & $\parid_4$ & $\parid_5$ & $\parid_6$  \\ \hline
	    	\rule{0pt}{1pt} \raisebox{-1.5mm}[-2mm][-2mm]{$\bar{S}^{\yff}$} \, & $9 \times 10^{-4}$ & $9 \times 10^{-12}$ & \textbf{0.520} & \textbf{1} & 0.005 & \textbf{0.217} \\ \hline
            \rule{0pt}{1pt} \raisebox{-1.5mm}[-2mm][-2mm]{$\bar{S}^{y_1}$} \,      & 0.001              & $6 \times 10^{-12}$ & \textbf{0.478} & \textbf{1} & 0.003 & $4 \times 10^{-4}$ \\ \hline  
        \end{tabular}
    \end{center}
\end{table} 

\section{Conclusion}
This paper presents a feedforward control scheme for electromechanical systems in which the output to be controlled cannot be directly measured or estimated. The method avoids real-time feedback by employing a run-to-run adaptation law that tracks a reference using the predicted error of an auxiliary output, requiring only an identifiable flat model.

This new control scheme, applied to the problem of soft landing in electromechanical switching devices, achieves results in terms of impact velocity reduction that are comparable to those of strategies that directly optimize that velocity. The main advantage over previous approaches lies in its ability to accurately track the reference trajectory.
Furthermore, a sensitivity analysis confirms that trajectory tracking can be achieved even under conditions of insufficient system excitation.

Future work will extend the feedback loop to exploit the full time-domain prediction-error signal, rather than relying solely on its integral, to potentially improve both tracking and parameter estimation.

\section*{DECLARATION OF GENERATIVE AI AND AI-ASSISTED TECHNOLOGIES IN THE WRITING PROCESS}
During the preparation of this work the authors used ChatGPT in order to improve the clarity and quality of the language. After using this tool, the author reviewed and edited the content as needed and take full responsibility for the content of the publication.

\end{document}